\documentclass[12pt]{article}
\usepackage{graphicx}
\begin{document}
\title{Towards the Intensity Interferometry Stellar Imaging System}
\author{M. Daniel, W.J. de Wit, D. Dravins, D. Kieda, \\
S. LeBohec,  P. Nunez, E. Ribak \\
for the Stellar Intensity Interferometry working group \\(within IAU commission 54)}
\maketitle
\noindent 
{\bf Contact}: David Kieda, University of Utah, Phone: 801 581 6901,\\
Email: kieda@physics.utah.edu
\begin{abstract} 
The imminent availability of large arrays of large light collectors
deployed to exploit  atmospheric Cherenkov radiation for gamma-ray astronomy
at more than $\rm 100\,GeV$, motivates the growing interest in application
of intensity interferometry in astronomy. Indeed, planned arrays numbering up to
one hundred telescopes will offer close to 5,000 baselines, ranging from less
than 50\,m to more than 1000\,m. Recent and continuing signal processing 
technology developments reinforce this interest. Revisiting Stellar Intensity Interferometry
for imaging is well motivated scientifically. It will fill the short
wavelength (B/V bands) and high angular resolution ($\rm<0.1\,mas$) gap left open by amplitude
interferometers. It would also constitute a first
and important step toward exploiting quantum optics for astronomical
observations, thus leading the way for future observatories. In this paper we outline science cases,
technical approaches and schedule for an intensity interferometer to be
constructed and operated in the visible using gamma-ray astronomy Air
Cherenkov Telescopes as receivers.      
\end{abstract} 
\vfill\eject

\section{Key science goals}
Stellar astrophysics distinguishes itself by the great progress that has been made in understanding the origin and evolution of the basic astrophysical phenomena, despite a physical inability to directly image
the stellar surfaces. Nearby stars have maximum angular extent of tens of milli-arcseconds (mas); direct imaging of objects in this angular range requires use of optical interferometry with baselines of one hundred meters or more. Until now, only a handful of stars (in addition to the Sun) have  been directly imaged to some extent using interferometric techniques. 

Of numerous interferometric methods, two are especially well suited for
imaging of stellar surfaces.  Stellar amplitude (Michelson) interferometry
(SAI) was employed (very simply) in the 1920's to measure a few cool stars,
but the technique was strongly control-technology-limited.  In the 1960's,
stellar intensity interferometry (SII) was employed to measure some 32
 stars \cite{brown1974a}, but it was, at that time, very sensitivity-limited.
More recently, SAI has been extensively developed, with several arrays
offering telescope apertures in the range 0.5-10\,meters and baselines up to
$\sim300$\,meters.  Modern amplitude interferometry has achieved sub-mas
resolution, with many applications to stellar physics \footnote{See the numerous Astro2010 Science White papers http://usic.wikispaces.com/ASTRO2010+White+Papers}

Amplitude interferometry as currently deployed, however, has significant
limitations.  In order to image hot stars (typical diameters $\rm < 1\,mas$) SAI would need considerably longer baselines, and would need to operate in the visible (difficult) or blue (very difficult). Intensity interferometry is well matched to blue/visible operation, and scales to long baselines with ease (the atmospheric stability requirements are a factor of a million less stringent for SII than for SAI). 

There are excellent technological motivations for renewing interest in
SII astronomy. First, the tremendous progress of signal processing technology
since the 1960's can improve sensitivity and makes implementation more cost
effective and reliable. This opens up the possibility to simultaneously
exploit a very large number of baselines to achieve a dense sampling of the
interferometric plane (in which the Fourier transform of the image is to be
measured). Second, within the next decade, extended  arrays of large diameter
light collectors are being planned for the next generation of Atmospheric
Cerenkov Telescope (IACT) Observatories, such as CTA \cite{CTA} and AGIS \cite{AGIS}. These projects will involve up to a hundred
telescopes distributed over several km$^2$ area. Such dense sampling
of an optical signal over the large area can sample the interferometric u-v
plane with close to 5,000 baselines ranging from a few tens of meters to more
than a kilometer. The combination of the dense sampling as well as the shorter
wavelength measurements can allow the  u-v interferometric plane coverage to
potentially measure stellar diameters down to the 0.1\,mas domain, a
capability that will open up a new field of ultra-high precision ground-based
astronomical imaging.  SII would offer a very complementary functionality to a
Cherenkov array, as the SII operation, on relatively bright stars, would be
far less sensitive to sky brightness, and could operate under moonlight, 
during times not useful for Cherenkov studies.

We envision a two-stage development and implementation of modern SII.  In the first stage, SII could be implemented on one or more existing Cherenkov arrays, immediately offering the advantages of blue operation and high resolution.  In a second stage, SII could be implemented as an augmentation of a next generation Cherenkov array, with an extremely rich range of baselines providing very high resolution and imaging capability unlikely to be matched by SAI in the foreseeable future, and in particular opening the possibility of very detailed imaging of young stars.

The ongoing  developments  in this
direction are presented in the following sections after an outline of the science potential offered by the proposed
re-deployment of SII. The capability to address various questions depends on the actual design of a modern intensity interferometer.  In the following sub-sections, in order to discuss the value of a few selected science topics which could be advantageously investigated, we adopt a conservative limiting visual magnitude $\rm m_{v}<8$ and a resolution of  0.1\,mas. 

\subsection{Pre-main sequence stars (PMS)}
Pre-main sequence stars  are young stars that are contracting
towards the main-sequence, but are powered only by gravitational collapse: 
 core temperature and pressure are insufficient to ignite hydrogen fusion. 
 Key questions relating to the  physics of mass accretion
and PMS evolution can be addressed by means of very high resolution
imaging.  Spatially resolved studies will address the absolute
calibration of PMS tracks, the mass accretion process, continuum
emission variability, and stellar magnetic activity, including the 
formation of potential jet emission. 

Features on  the stellar surface may also be directly imaged.   Hot spots on
the stellar surface deliver direct information regarding rotation via the von Zeippel effect, as 
well as tracing the accretion of material onto 
the stellar surface. Cool spots, on the other hand, may cover 50\% of
the stellar surface; they may be produced by magnetic field effects (sun-spots), or
as a product of the slowly decaying rapid rotation of young stars. Measurement of
the number, distribution, and then variations of hot and cool spots will constrain 
the interplay of rotation, convection, and chromospheric
activity. It may also provide direct practical application
for exploring the anomalous photometry  observed in young stars\cite{Stauffer2003}.
In addition, the presence of hot and cold features on the stellar surface is 
not only limited to PMS stars; observation of such features on main-sequence stars may allow
detailed study of sun-like transient phenomena on a wider class of main sequence
stars. Because the most luminous stars will provide the strongest 
interferometric signal, U/V band measurements of stellar surface features  by SII
can provide optimal contrast.

In the last decade, several young, coeval stellar groups have been
discovered in close proximity ($\sim$50\,pc) to the sun. Prominent examples of
nearby coeval stellar groups include the TW Hydra and $\beta$ Pic co-moving groups.  
The proximity of the co-moving groups ensures  that their members  are bright.
The majority of the spectral types within these nearby groups
range between A and G-type, and about 50 young stars have $\rm
m_{v}<8$.  Their close proximity renders the co-moving groups relatively 
sparse making them very suitable, un-confused targets. The relative sparseness also 
must certainly generate incomplete group membership. As stellar distances continue to
be refined, additional members of these groups may be discovered,  making it likely 
that the known number of young, bright stars will continue to
increase with time. The ages  of the stellar coeval groups lie in the range between 8
to 50\,Myr (see \cite{Zuckerman2004} for an overview).  These young ages  imply that a substantial fraction of the low-mass members
 are still in their PMS contraction phase.

The evolutionary PMS tracks in the Hertzsprung-Russell diagram are usually
calibrated using spectroscopic binaries, which do not allow to
 resolve the individual stellar masses. Statistical methods are used to estimate
the inclination angle of the systems and estimate the average masses for an ensemble
of stars. By directly imaging the binary orbital parameters, the inclination 
of the system can be determined, leading to direct measurement of the 
masses of the binary components. 
This is a fundamental exercise not only for PMS stars but for 
all spectroscopic binaries in any evolutionary stage. Measurement
 of angular sizes of individual PMS stars in combination with a 
distance estimate (e.g. GAIA) allows a direct comparison between
 predicted  and observed sizes of these gravitationally contracting 
star on a star-by-star basis, rather than on an average basis over an 
ensemble of stars. 

\subsection{Cepheid distance scale}
Measuring diameters of Cepheids is a basic method with far reaching
implications for determination of the distances of nearby galaxies. 
Currently, optimum use of a Cepheid distance scale requires calibration of
distances, and correction for Cepheid atmospheric factors.  Hipparcos (and in
the future GAIA and perhaps SIM) provide precision distances for
calibration. SII promises much higher resolution and extension of Cepheid
calibration to greater distances with the Baade-Wesselink method
\cite{sasselov1994}. In an exciting development, intensity
interferometry was recently proposed as an approach for the measurement of
absolute distances by exploiting the time curvature of the observed light wave
fronts \cite{ralston2008}.  Meanwhile, SAI is already providing critical
ancillary information, such as the atmospheric p factor, for Cepheids of known distance \cite{merand2005}.

\subsection{Fast rotating (Be) stars}
Classical Be stars are particularly well-known for their
close to break-up rotational velocities as deduced from photospheric
absorption lines.  In addition, they show both IR excess and Balmer line emission due to a gaseous circum-stellar disk.  The line emission appears and disappears on timescales of months to years.   Photometric observations of Be star disks sometimes provide evidence for evolution of the disk structure into several ring structures,  before  the gas disappears into the interstellar medium (e.g. \cite{dewit2006}). 
The Be-phenomenon is fairly common (fraction of Be stars to normal B-type peaks at nearly 50\% for B0 stars, \cite{zorec1997}), and is therefore indicative of a fundamental stellar physics phenomenon. There are about 300 Be stars\footnote{\tt http://www.astrosurf.com/buil/us/becat.htm}
brighter than $\rm m_{v}=8$, roughly corresponding to a distance
limit of 700\,pc. 

The disk structure, fast rotation, and emission line phenomena in Be stars appear to be related. Although the exact underlying Be mechanism has yet to be identified, the Be phenomenon is probably related to the high rotational velocity of the star. Absorption line studies cannot provide an accurate rotational velocity in these objects due to strong gravity darkening at the equator and brightening at the pole areas. 
However, direct measurement of the physical shape of the rotating star can provide a direct measure of the rotational speed (see $\alpha$\,Eri with the VLTI,
\cite{desouza2003}).  The peak blackbody wavelengths of the Be spectra are in
the U/V bands. Since ground-based Michelson interferometry is still restricted to
IR wavelengths for imaging, the long-baseline   
coverage of the interferometric u-v plane is reduced , even for the same
physical telescope baseline. In addition, the lower photon flux in the IR tail
further decreases the number of observable Be stars. 

Since SII is  substantially less sensitive to atmospheric
turbulence than Michelson interferometry, observations of Be stars in the U/V
band can enable high resolution studies of Be stars over substantially longer
baselines. A rough estimate of the effective temperatures of the objects in
the Bright Star Catalog reveals there are approximately 2600 stars brighter
than visual magnitude 7 and hotter than 9000\,K. Their typical angular sizes
range between 0.5\,mas and 5\,mas making them observable even with a fairly modest scale intensity interferometer array. 

\subsection{Other Stellar Physics}

Optical interferometry is still in its infancy, but already more than 150 science publications show the dramatic return from high angular resolution studies of stars, ranging from YSO's and pre-planetary disks, to MS stars, debris disks, multiple stars, mass loss in all kinds of systems, and PMS evolution.  SII promises to complement and extend such programs to blue/visible imagery with much higher resolution, and at comparatively modest incremental cost.  Rather than develop specific science topics further here, we refer to the typically 4-5 SAI publications currently appearing in the literature each month.

\section{Technical overview}

\subsection{Intensity Interferometry Technique}
The intensity interferometry technique does not rely on the 
actual interference of light rays as in Michelson interferometry. Instead, the
interferometric signal, the degree of mutual coherence, is characterized 
by the degree of correlation of light intensity fluctuations observed at two 
different detectors.  In practice, the degree of mutual coherence is measured 
using fast temporal correlations between narrow optical waveband 
intensity fluctuations observed by  two (or more) telescopes separated by a baseline distance. 

The principal component of the intensity fluctuation is the classical shot
noise which will not demonstrate any correlation between the two separated
telescopes. The intensity interferometric signal is  related to a smaller noise component:
the wave noise. The wave noise can be understood as the 'beat frequency' in 
optical intensity between the different Fourier components of the light 
reaching the telescopes.  This wave noise will show correlation between  
the two detectors, provided there is some degree of mutual coherence
between the light received at the two telescopes.  
As per equation \ref{coherence}, this intensity correlation 
(the time integrated product of the intensity fluctuations $\rm \Delta i_1$ and 
$\rm \Delta i_2$ in two telescopes) is positive. The correlation 
provides a measurement of the square of the degree of coherence of the light
 at the two detectors $\rm |\gamma(d)|^2$  (the fringe visibility in a Michelson
interferometer) where $\rm d$ represents the distance separating the 
telescopes. \cite{labeyrie2006}. 
\begin{equation}
|\gamma(d)|^2 = {{<\Delta i_1 \cdot \Delta i_2>}\over{<i_1><i_2>}}
\label{coherence}
\end{equation}
 
According to the van Cittert-Zernike theorem,  the
complex degree of coherence ($\rm \gamma(d)$)  is the normalized Fourier transform of the source
intensity angular distribution. In the case of a source that can be modeled as a uniform
disk with angular diameter $\theta$, the various baselines used in the observation
sample an Airy function on the ground.  As a consequence, when the baseline
$\rm d$ is too small for resolving the observed object, $\rm |\gamma(d)|^2=1$ and
$\rm \Delta i_1$ and $\rm \Delta i_2$ are maximally correlated. Conversely, when the 
telescope baseline is sufficient to resolve the Airy disk, $\rm |\gamma(d)|^2<1$, 
and $\rm \Delta i_1$ and $\rm \Delta i_1$ are still correlated to some extent. For 
observations at wavelength $\lambda$, the degree of coherence $\rm \gamma(d)=0$ 
when $\rm d = 1.22 \lambda / \theta$. An intensity interferometer with 1\,km 
baseline operating at a 400\,nm wavelength should therefore be able to probe
features with angular extent less than  0.1\,mas.  

\begin{figure}
\centerline{\includegraphics[width=9cm]{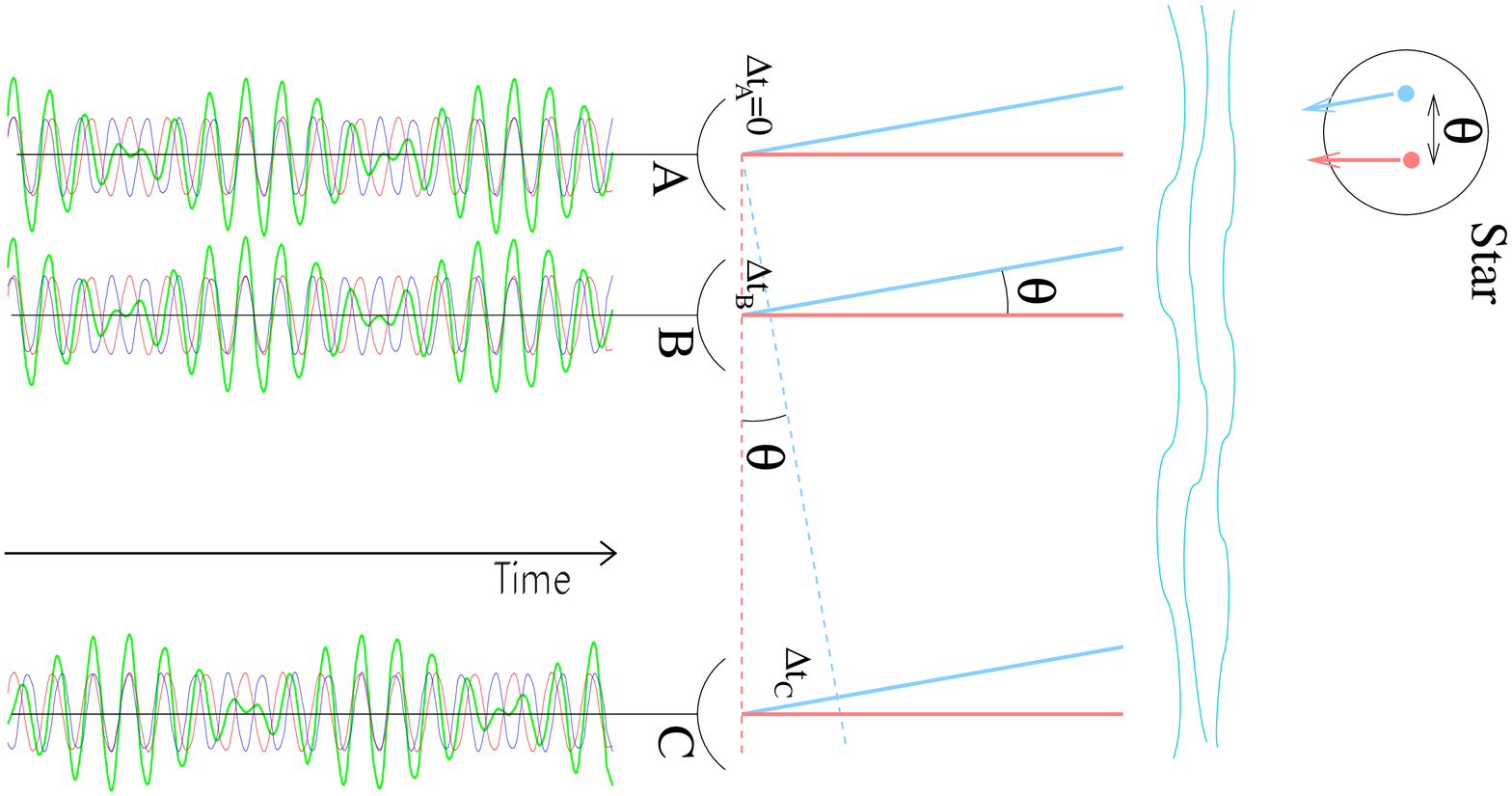}}
  \caption{Considering two harmonic emitting points in the source, with
  slightly different frequencies, telescopes A and B being close together
  receive beats almost at the same time. This corresponds to a high degree of
  correlation. Telescopes A and C being further apart do not receive beats in
  phase and the degree of correlation is then lower \cite{lebohec2008b}.}
  \label{principle}
\end{figure}

The important point is that the technique relies on the  correlation between 
the (relatively) low-frequency intensity fluctuations between different detectors, and does not rely on the relative phase of optical waves at the different detectors 
(see Figure \ref{principle}). 
The requirements for the mechanical and optical
tolerances of an intensity interferometer are therefore much less stringent
than in the case of a Michelson interferometer. One strong advantage of intensity 
interferometry is therefore its mechanical and atmospheric
robustness: the required accuracy depends on the electrical bandwidth of the
detectors, and not on the wavelength of the light. This opto-mechanical
robustness also means that the atmosphere does not influence the performance
of the instrument even in the U and V bands. 

The major drawback of SII is that copious quantities of light are necessary to
observe the wave noise signal in the presence of the much larger shot noise.
Very large light collectors  ($50-100$ m$^2$ area) are necessary to have sufficient
statistical strength to observe non-Poisson fluctuations around the mean photon
intensity. However, the tolerance of
SII to path length differences makes such light collectors relatively
inexpensive. An intensity interferometer with 1\,GHz signal bandwidth, does
not need the optical surface of the collectors to be much more accurate than
3\,cm and the pointing does not need to be controlled to better than a few arc
minutes for 10\,m telescopes. Interestingly, these primary design requirements
are well matched by Imaging Atmospheric Cherenkov telescopes used for
gamma-ray astronomy above $\rm 100\,GeV$.  SII also appears to be the only  
viable
tool for high resolution studies through the atmosphere in blue wavelengths. 
Observations at short wavelengths will provide a higher 
contrast for measurement of spatial differences in the envelope of 
hot stars. These wavelengths 
can potentially extend imaging capabilities of bright stars to larger 
distances than Michelson interferometry.

\begin{figure}
\centerline{  
\includegraphics[width=8cm]{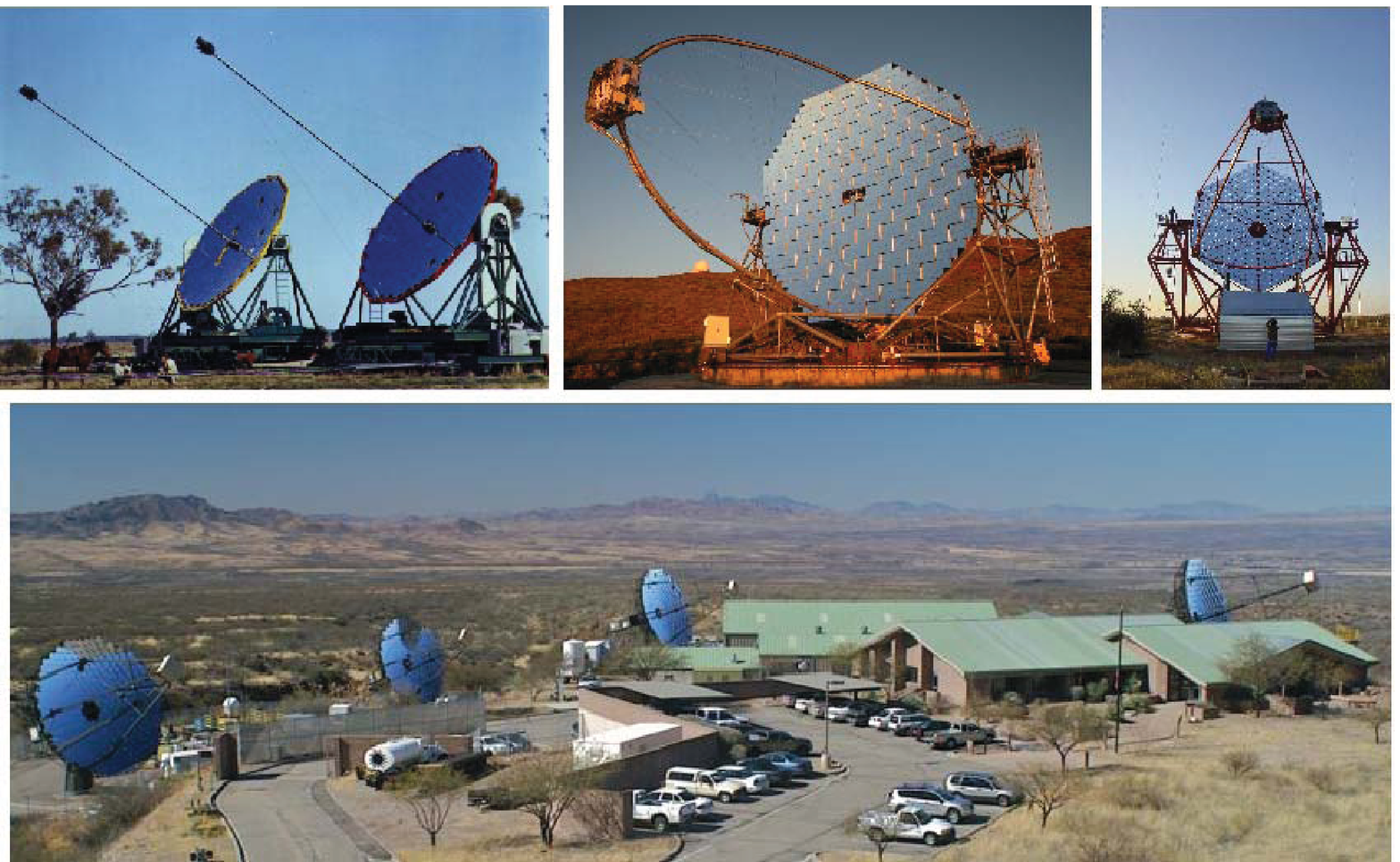}
  \includegraphics[height=5cm]{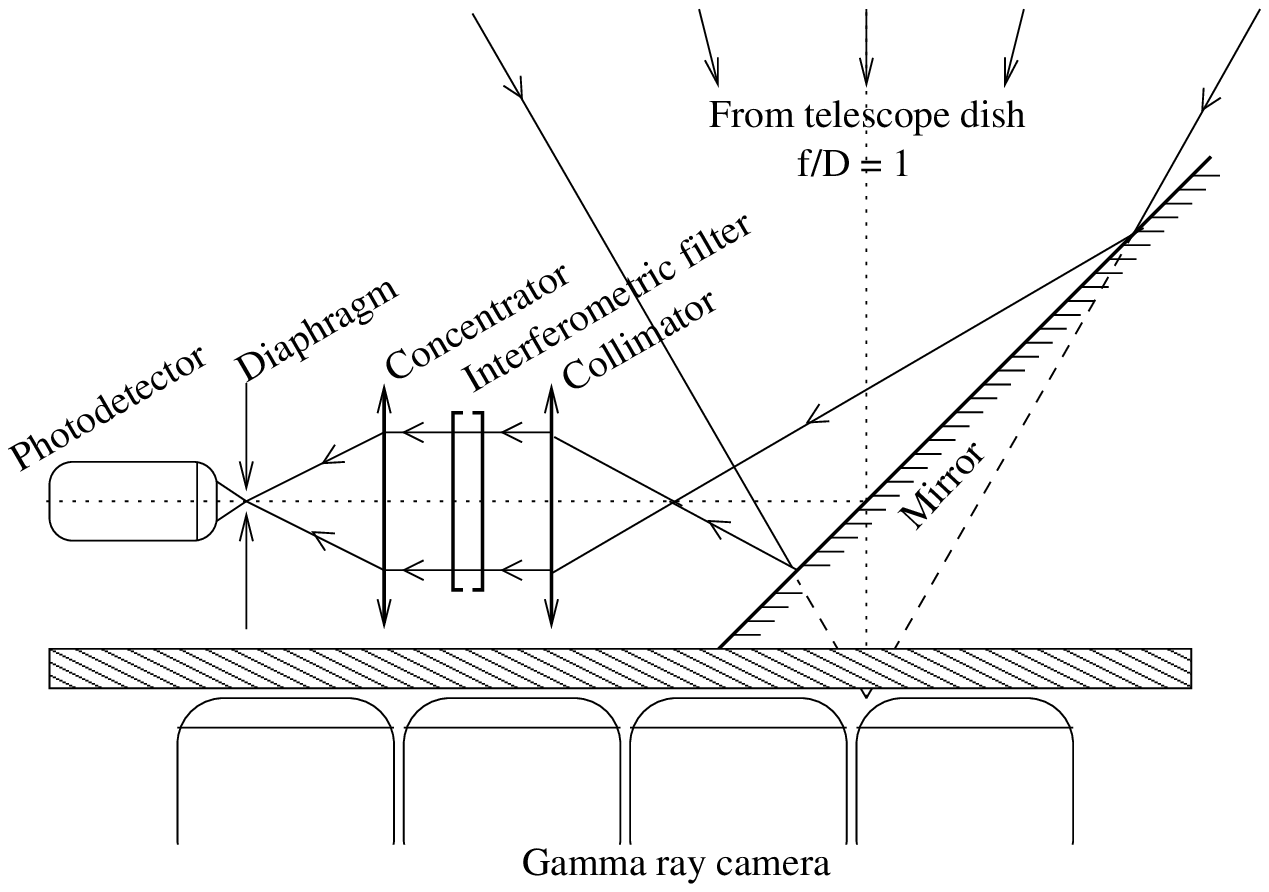}
}
  \caption{The two 6.5\,m Narrabri Stellar Intensity Interferometer telescopes 
  (top left) have many common points with Imaging Air Cherenkov Telescopes 
  such as the two 17\,m MAGIC telescopes (center top), the four 12\,m
  H.E.S.S. telescopes (right) or the four 12\,m VERITAS telescopes
  (bottom). On the right is a schematic of a possible implementation of SII on
  a Cherenkov telescope \cite{lebohec2006}. The light from the mirrors is reflected side way toward a
  collimator and analyzer before being focused on the photo-detectors. Such a
  simple system can be mounted on the shutter of the atmospheric Cherenkov camera so it does
  not block the view for gamma-ray observations.}
  \label{nsii}
\end{figure}

\subsection{Using Ground-based VHE gamma-ray Telescope arrays for SII }

The intensity interferometry technique was initially developed for radio
astronomy \cite{brown1952}. 
Hanbury Brown and Twiss subsequently demonstrated that the technique could  also be employed at visible
wavelengths \cite{brown1957a, brown1957b}. Based upon the success of these initial measurements, 
the Narrabri Stellar Intensity Interferometer (NSII) was designed and
constructed \cite{brown1974}. 
NSII was located near Narrabri, Australia, and operated from 1965 to 1976 (Figure \ref{nsii}). 
The NSII  consisted of two telescopes, each 6.5\,m in diameter, with an 
11\,m focal length. The two telescopes were carried on trucks running on a 
circular railway track  188\,m in diameter. This allowed the interferometer to operate with a baseline 
of 10\,m to  188\,m. 
The combination of mirror area,
electronic bandwidth, and electronic noise  restricted observations to stars 
brighter than B = +2.5. Still, the 
NSII was the first instrument to allow the successful measurement of  angular 
diameters of main sequence stars. A total of 32 angular diameters were 
measured with the NSII, some as small as 0.4\,mas \cite{brown1974a}.  

Imaging Air Cherenkov Telescopes (IACT) are used for gamma-ray astronomy at
Very High Energies (VHE, energies greater than 100\,GeV). The IACT technique
relies on the fact that VHE particles and gamma-rays initiate
extensive air showers of high energy secondary particles in the atmosphere. 
Charged shower particles with sufficient kinetic energy will radiate optical
Cherenkov light into the atmosphere. This Cherenkov light at ground level is
strongly peaked at blue wavelengths, has a duration of only a few nanoseconds,
and is also very faint ($\rm \sim 10\,photons/m^2$ at 100\,GeV). Large
($>10\,m$ diameter) light  collectors with excellent U/V band reflectivity 
equipped with fast electronics are employed to detect this Cherenkov light.  
IACTs are typically used in widely-spaced arrays of 2 to 4 telescopes with
typically 100\,m inter-telescope distances in order to record stereoscopic
views of each shower. This telescope separation is chosen to match the extent
of the Cherenkov light pool at ground level \cite{weekes2003}, and is
dependent on the altitude of the  IACT observatory, Figure \ref{nsii} presents
examples of current IACT telescopes.  

Because the design requirements of IACT arrays and SII arrays are generally
similar, these facilities can potentially be used for both types of
astronomical observations \cite{lebohec2006,lebohec2008a}. Historically,  the 
two SII telescopes of the 
Narrabri interferometer were subsequently instrumented as  air Cherenkov
telescopes in a search for astronomical sources of very high energy ($\rm
E>300\,GeV$) gamma-rays \cite{grindlay1975a, grindlay1975b}.  

An existing IACT telescope can be converted into a SII receiver by 
adding an optical collimator, an interferometric filter,  a photo-detector 
and front end electronics on the focal plane of the IACT telescope. The 
SII instrumentation could be mounted in front Cherenkov cameras as illustrated on the right side of Figure \ref{nsii}.
It is desirable to have the SII instrument package integrated
into the camera shutter so that it does not interfere with normal gamma-ray observations.
The HESS telescopes have already prototyped such a mounting system for performing 
photo-metric measurements of stars with a photomultiplier tube \cite{deil2007}. 
The SII instrument will send the SII signal from each telescope to a central facility to
be combined and correlated over various baselines, thereby 
filling out the u-v interferometric plane. 
The length and orientation of the effective baseline between two IACT
telescopes depends on the position of the star in the sky.
 At the time of the final data analysis, the interferometric (u,v) plane is binned,
taking into account the time-varying baselines from each telescope pair.

Although IACT telescopes can be used for SII, IACT arrays are optimized for 
high energy gamma-ray astronomy. Optimizations for highest gamma-ray sensitivity 
with lowest cost  will constrain the  SII capabilities of these instruments. 
For example, the signal to noise ratio in an SII instrument depends upon the 
square root of the 
correlator bandwidth. IACT telescope designs  typically employ a large ($>3^\circ$ diameter) field of view
in order to capture the entire length of the optical Cherenkov signal as well as to 
efficiently map the night sky for new sources of VHE gamma-rays. 
The requirement to minimize optical aberration over the wide field of view has led
several groups to adopt the Davies-Cotton \cite{daviescotton1957} telescope design. 
However, this design does not preserve iso-chronicity (simultaneous arrival at the focal plane of all photons collected by single flash of light across the full mirror surface). 
For 12\,m telescopes such as in HESS and VERITAS 
\cite{veritas2008}, the telescope optics spreads an optical light impulse out to a duration of several nanoseconds at the focal plane. This effectively convolutes any optical signal with an effective
filter bandwidth  of $\rm\sim100\,MHz$. With parabolic telescopes such as in MAGIC 
(17\,m) or H.E.S.S. II (29\,m), the effective signal bandwidth filter is on the order of 1\,GHz. 

The optical quality of the IACT telescope will also constrain the limiting visual magnitude for
SII applications. The optical point spread function of typical Davies-Cotton IACT design is 
$0.05^\circ$. This point spread function is well matched to the typical angular width of an
air Cherenkov light image ($0.1^\circ$). However, because of the corresponding integration of night sky
background light, these telescopes could not be used for SII observations of
stars any fainter than visual magnitude $\sim 9.6$ \cite{lebohec2006} . 

In contrast with movable Narrabri SII telescopes, the telescopes in an IACT array
are at fixed positions on the ground. The SII signals from different telescopes
will have to be brought back in time coincidence for the correlation to be measured;
the time delay between arrival of the optical telescopes depends upon their
projected path length difference to the star. The path length difference changes as the
star transits through the night sky. Electronic signal delays 
will be added to each photo-detector signal  to correct the arrival time, and thereby restore temporal coincidence between intensity fluctuations. The minimum signal delay accuracy 
is determined by the optical bandwidth of the telescope.
To preserve the highest frequency SII correlations between two telescopes with
an accuracy better than 10\%, the signal time delays must be controlled to an
accuracy better than 7\% of smallest signal period. For example, a $\rm \sim
100\,MHz$ system requires time delay correction accuracy of $\rm \sim 0.7\,ns$.
An analog delay system with similar requirements has been
operated by the CELESTE experiment \cite{pare2002}.

IACT gamma-ray observations are generally limited to dark, moonless nights. 
During periods when the moon is in crescent phase, IACT telescopes may operate with reduced
sensitivity, or they may use this time to perform optical calibrations and alignments. 
For VERITAS, HESS, and MAGIC,  the equivalent of $\sim10$ nights each month during near-full
moonlight are unused for gamma-ray astronomy, and could be made available for
SII observations. This corresponds to $\sim 800$ hours of SII observation per
year.  Assured telescope access of this magnitude is more than sufficient for development of a vigorous SII science program. At the same time, the SII program will have minimal impact on existing VHE gamma-ray science programs. If the SII
science program is successful using the freely available 800 hours/year on the next generation IACT arrays, it would be advantageous to consider building a
future dedicated SII facility with a square kilometer array on varying, non-periodic baselines. The first-light time-frame for a dedicated facility would be in the 2020-2030 decade, or later. 

\begin{figure}
\centerline{
  \includegraphics[width=4.5cm]{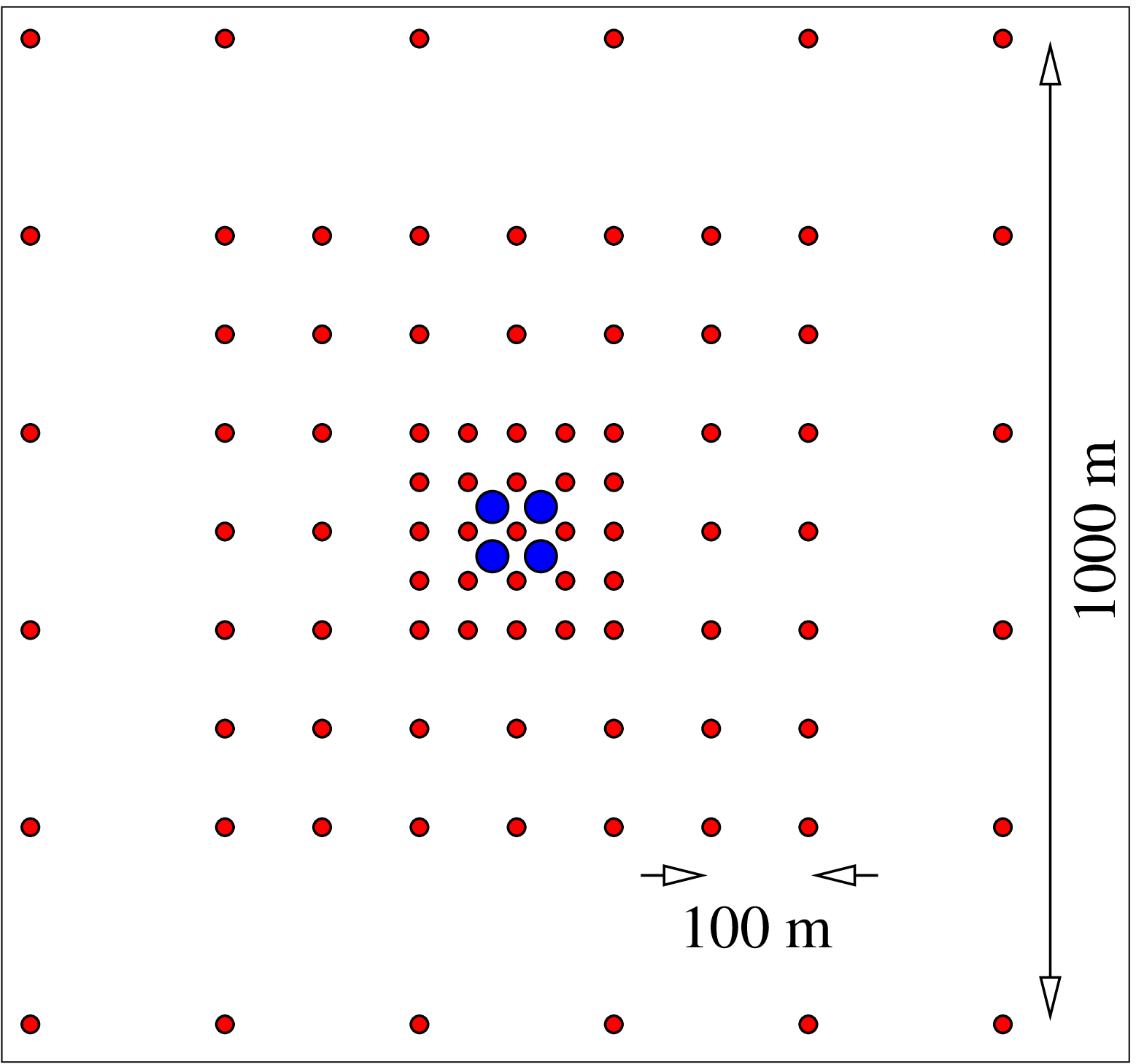}
  \includegraphics[width=4.5cm]{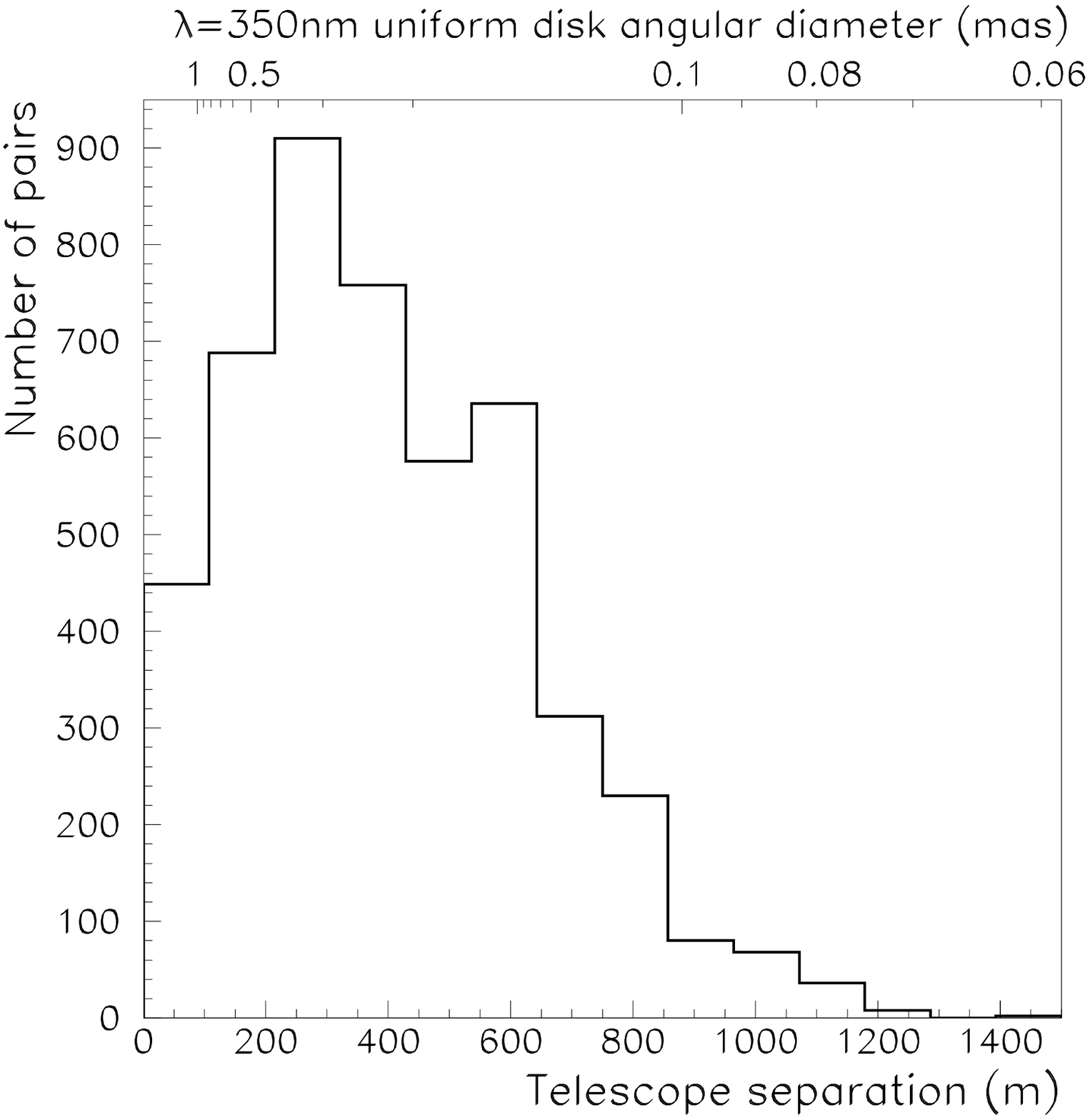}
  \includegraphics[width=4.5cm]{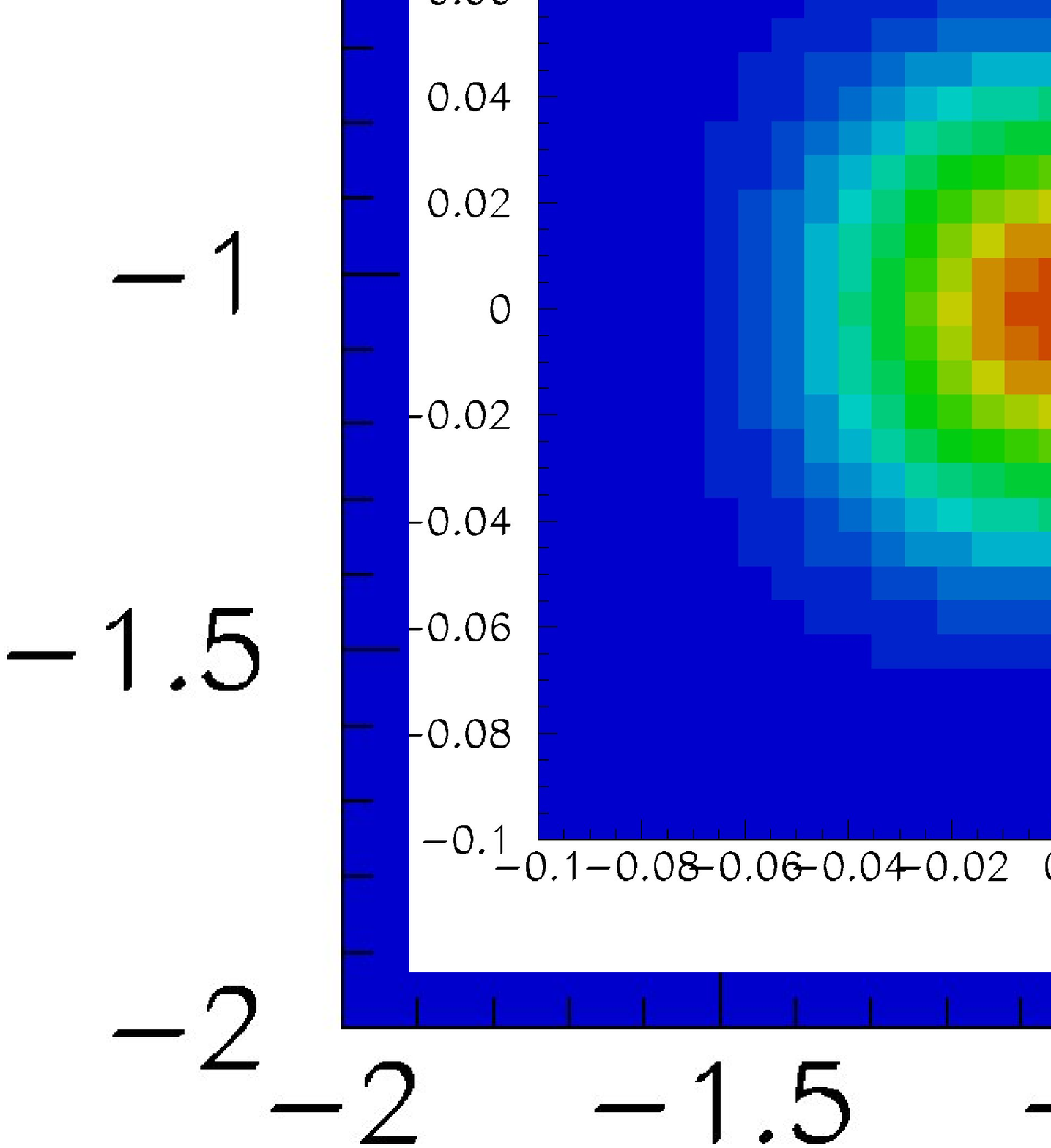}
}
  \caption{The proposed lay-out for the future CTA (left) includes 85 $\rm
    100\,m^2$ dishes (red dots) and four $\rm 600\,m^2$ dishes (blue dots). It
    would offer 3,916 baselines \cite{lebohec2008a} ranging from 35\,m to 
1,414\,m (center) and
    allowing imaging at 400\,nm over an aliasing free 1.6\,mas field of view with
    a better than 0.1\,mas angular resolution (right).}
  \label{cta}
\end{figure}

\subsection{Capabilities of an SII implemented within the CTA/AGIS IACT arrays}

Next generation IACT arrays such as  CTA \cite{CTA} and AGIS \cite{AGIS} are
planned for construction and operation during the decade 2010-2020. 
These square kilometer arrays  will use up to 100 telescopes (Figure \ref{cta}) providing up to 5000 base-lines ranging from $\sim 50$\,m to $\sim 1.4$\,km. 
Calculations indicate that a CTA-type IACT array 
used as an SII observatory would allow direct imaging on 
angular scales between several mas to less than 0.1\,mas. 

\begin{figure}
\centerline{
  \includegraphics[width=6cm]{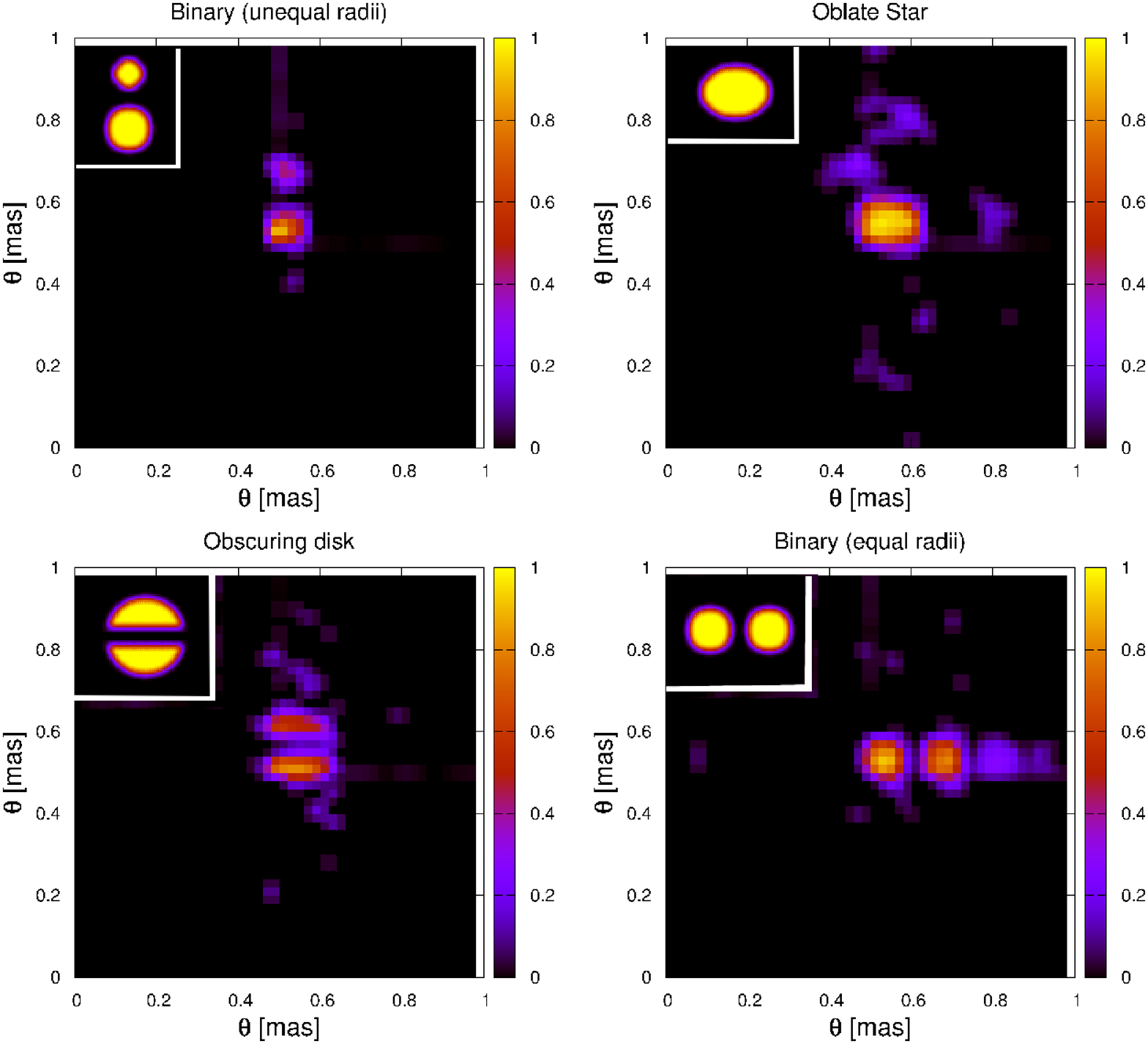}
  \includegraphics[width=6cm]{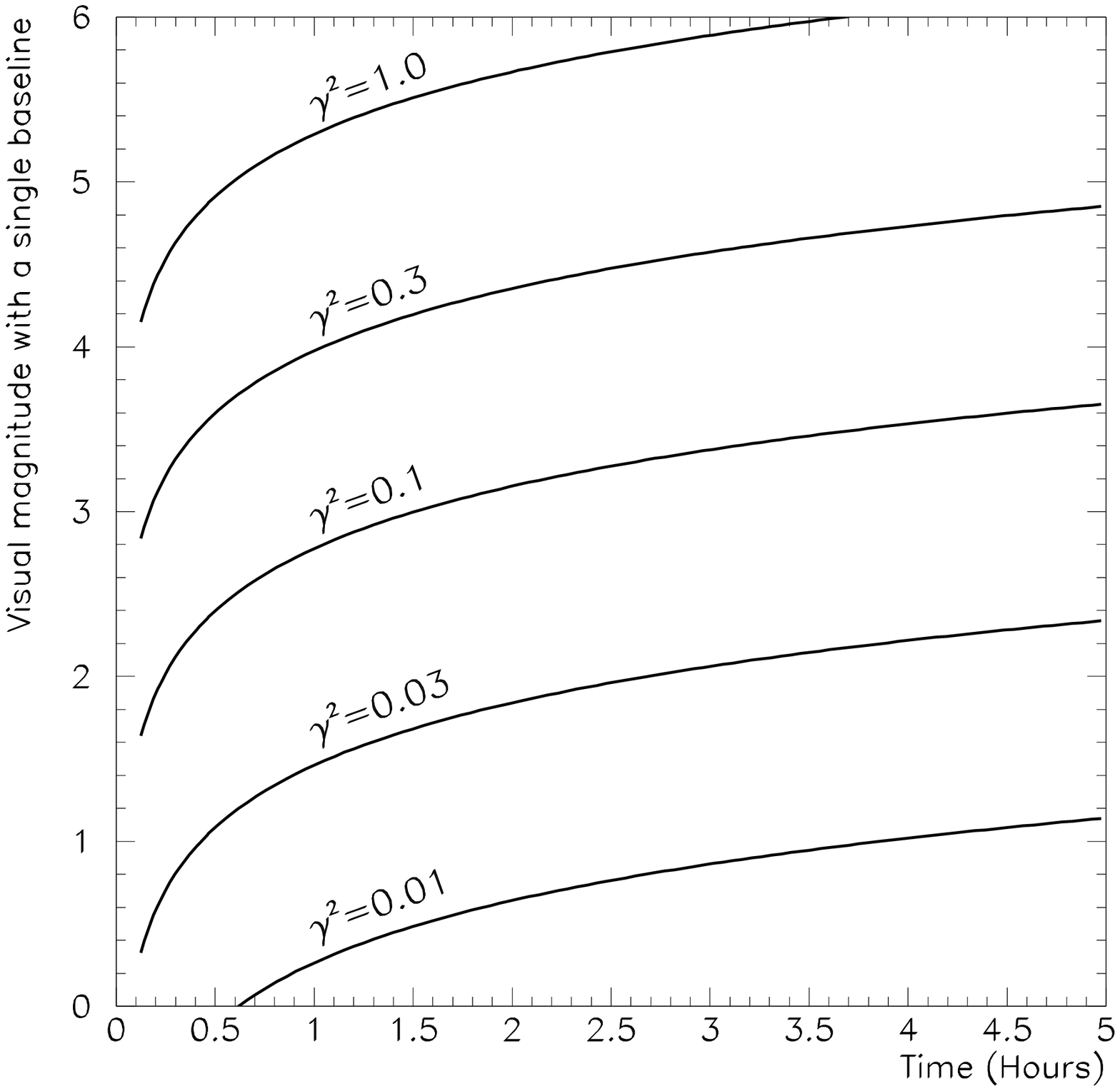}
}
  \caption{One the left, four examples of reconstructed images from a 
simulated SII observation
using an IACT telescope array. The simulation uses a wavelength of $\sim 400$\,nm with
one hundred CTA-like  telescopes and a inter-telescope separation of $\sim 100$\,m.
 The pristine image is shown at the top left in each example. The images were 
produced \cite{nunez} using an algorithm based on the Cauchy-Riemann equations
\cite{holmes2004}.  The analysis does not yet include a realistic 
noise component, which is still actively being investigated. One the right, for a range of degrees of coherence $| \gamma |^2$, the
  visual magnitude for which a five standard deviation measurement is possible 
  with a single baseline
  is indicated as a function of the observing time. We assumed $\rm A=100\,m^2$
  light collectors with quantum efficiency $\alpha=30\%$ and a signal
  bandwidth of $\rm 100\,MHz$. Baseline multiplicities of 10, 30, 100 and 300 improve
  the sensitivity by 1.25, 1.85, 2.51 and 3.11 magnitudes respectively. This is for
  model independent imaging only. Model fitting would allow the study of
  objects with fainter visual magnitudes. }
  \label{paul}
\end{figure}

The SII technique provides a measurement of the magnitude of the Fourier 
transform of the image, but does not directly measure the phase information 
of the Fourier transform. Several phase reconstruction techniques have been proposed which employ higher order intensity correlations to recover the phase information
\cite{gamo1963, fontana1983, marathay1994, vildanov1998}. The exploitation of
higher order correlations has implication on the design of the Intensity
Interferometer. Signals from telescope triplets, quadruplets and more have to
be brought together to be combined.  
A possibly more promising  method exploits the analyticity properties of the 
Fourier 
transform to establish a relation between finite differences 
in the magnitude and in the phase of the Fourier transform 
\cite{holmes2004}. A typical implementation used to recover high resolution
source images with a CTA-like array is illustrated on the left of Figure 
\ref{paul}
 This method has been demonstrated to be very robust against 
potential noise contamination \cite{holmes2009}. 

The sensitivity of the SII is determined by the shot noise in the incoming photon stream from each telescope.  In the measurement of the degree of correlation of intensity fluctuations
 between two telescopes, the signal to noise ratio ($\rm S/N$) is 
\begin{equation}
({S / N})_{RMS} = A \cdot \alpha \cdot n(\lambda) | \gamma(d) |^2 \sqrt{\Delta f \cdot T /2}
\end{equation}
where $\rm A$ is the light collection area of one telescope, $\alpha$ is the
quantum efficiency, $\rm |\gamma(d)|^2$ the square of the magnitude of the degree
of coherence for a given baseline $\rm d$. The right side of Figure \ref{paul} shows the expected
$\rm S/N$ ratio as a function of the exposure time for conservative telescope
parameters ($\rm A=100\,m^2$, $\alpha=30\%$ and $\rm \Delta f=100\,MHz$). With five hours of observation, 
a single baseline can provide
measurements of $|\gamma|^2 = 0.3$ and 
$|\gamma|^2 = 0.03$ for stars of visual magnitude 4.8 and 2.4 respectively, with  a $5\sigma$ statistical significance. 

The statistical sensitivity of these observations can be further improved by 1 to 3 magnitudes (depending of number of baselines used and specific array design)  by exploiting the redundancy of the baseline measurements in the  
full array. This could be further improved if additional optical channels are 
available at each telescope. Finally, when the information sought does not 
require a model independent reconstruction of the image (as in the case of a 
binary, a fast rotating oblate star, a star with an obstructing disk, etc), 
all the baselines can be simultaneously exploited to push SII sensitivity to 
objects approaching $\rm m_v= 9$. 

For typical Cherenkov telescope quality (PSF $=0.05^\circ$, area $=100\,m^2$),
the ultimate sensitivity ( $\rm m_v= 9$) is dominated by the integrated night
sky background in the PSF, and not by the SII signal statistics. The
observation of objects fainter than $\rm m_v= 9$ requires both higher telescope optical quality as well as extended observation time. For each additional visual
magnitude,  the required observation times increases by a factor $> 6$. This
of  does not preclude the utilization of larger telescopes (Area $\rm >
100\,m^2$) as long as the source remains unresolved across the diameter of the
mirror. In particular, a five hour observation by a single pair of CTA $\rm 600\,m^2$ telescopes will provide $|\gamma|^2 = 0.3$ and $|\gamma|^2 = 0.03$ measurements
for visual magnitude 6.7 and 4.3 stars, respectively.

\section{Technology drivers}
The currently  envisioned SII implementation does not depend on the
development of any critical technological capability. There are technology developments, however, that can benefit the capabilities of future large scale Intensity Interferometer arrays. These technological developments may extend SII to higher sensitivities, thereby extending the imaging capabilities to dimmer visual magnitudes. In this section, we describe the potential impact of two  technological drivers.

\subsection{High speed data recording and handling}
A major improvement in the ultimate sensitivity of a large SII array can be accomplished by continuously recording each telescope's stream of intensity
fluctuations with minimal signal processing, and calculating the inter-telescope correlation functions post-observation.  This capability can  allow
studying the degree of correlation as a function of time-lag between the  two 
photon streams. Ofir and Ribak \cite{ofir2006} have recently proposed that the time-lag dependence of the correlation carries detailed information regarding the source emission process. The availability of the full data stream from each telescope can also allow analysis techniques based on higher order intensity correlations (i.e. 2, 3 and more points). These higher order correlations are employed for both phase reconstruction analysis and also  for absolute source distance measurements based on light wave front curvature \cite{ralston2008, ofir2006}. 

The ability to handle such large data rates has only recently become technically feasible. A continuous $200\,MHz$ digitization of the photo-detector signal with 4 bit accuracy  generates $360\,Gb$ per hour of operation of a single
CTA telescope, and about $280\,Tb$ per observation night for the  
full array. Using existing technology such as off the shelf 1.5\,Tb SATA disks (2009  price : \$150 each) in  a distributed, parallel
RAID array, it is feasible to archive the data in real time for next-day processing. The next day analysis pipeline would employ a farm of
distributed processors to generate the intensity correlations and invert
the Fourier plane to generate the source image for each observation. The data storage requirements are minimal once the SII data has been correlated and inverted (10-100 Gbyte/night maximum), and so after analysis, the RAID data  storage farm  would be overwritten by the next night's data. As hard-disk technology continues to advance, it may eventually be possible  to archive the entire raw data set each night for longer term study and re-analysis.

Technological challenges associated with a 1-day turnaround pipeline analysis are
of similar magnitude to other state-of-the-art telescope systems  such as  LOFAR, ALMA, and SKA. The most significant technological challenge to a large SII array is probably in establishment of sufficient bandwidth across the widely distributed array that will allow efficient transport all data to localized data storage nodes for collection and correlation. The data collection and processing scheme must be designed to provide sufficient   redundancy and robust operation so that it may continue to perform its analysis task even when   one or more telescope systems in the array have failed.

\subsection{High speed \& high quantum efficiency photo-detectors}
\begin{figure}
\begin{center}
  \includegraphics[width=7cm]{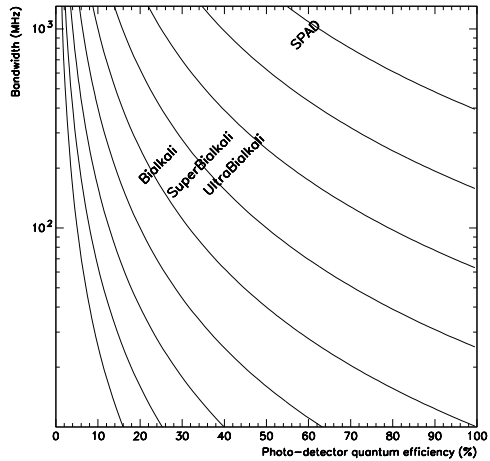}
\end{center}
  \caption{Sensitivity contour lines are separated by one visual magnitude in the
  signal bandwidth versus quantum efficiency plane characterizing
  photo-detectors. SPADs could give a 4 magnitude improvement compared to
  regular bialkali photomultipliers. However, the effective signal bandwidth
  if limited to
  100\,MHz for 10\,m Davis-Cotton f/1 Cherenkov telescopes. Larger Cherenkov
  telescopes (MAGIC, H.E.S.S. II) are parabolic and provide a signal bandwidth
  close to 1\,GHz.}
  \label{photosensit}
\end{figure}

As described previously, the $S/N$ ratio for SII is proportional to the photo-detector quantum efficiency $\alpha$  and also to the square root of the combined photo-detector/optical bandwidth $\Delta f$. For a fixed stellar magnitude, improvements in the $S/N$ ratio (and therefore visual magnitude sensitivity) can therefore be achieved through technological improvements to the photo-detector quantum efficiency and speed (Figure \ref{photosensit}).

A standard bialkali photomultiplier tube will provide quantum efficiency
$\alpha=25$\% at  optical wavelength $\rm \lambda=350\,nm$ and signal bandwidth
$\Delta f=200\,MHz$. Super-bialkali photomultiplier tubes have recently become
available with $\alpha=35$\%, and Ultra-bialkali with
$\alpha=43$\%. Consequently, one may improve the $S/N$ ratio by a factor of
two for a minimal cost per telescope. The sensitivity improvement is
equivalent to doubling the mirror area with a standard bialkali photomultiplier tube. The application of ultra-bialkali photomultipliers to SII is fairly straightforward and relatively inexpensive. This technology has yet to be deployed on a current SII instrument;
The SII observation environment poses special challenges due to the strong optical intensity at the focal plane. Issues regarding temperature stability, saturation, and photomultiplier tube lifetime have yet to be explored. This technological development is underway and should proceed rapidly over the next 
few years.

The highest quantum efficiency photo-detectors commercially available are
based upon semiconductor technology; these devices (e.g. avalanche
photodiodes) may approach $\alpha=95\%$ at $\rm \lambda=330\,nm$. Large area silicon devices suffer from substantial inherent electronic noise related to the silicon bulk resistivity and the depth of the depletion layer at the PN junction. In order to achieve single photon sensitivity, recent technology has focused on the development of large arrays of very small area Single Photon Avalanche Photodiodes  (SPADs). By minimizing the SPAD area (and hence the capacitance), each SPAD achieves sufficiently low inherent electrical noise to allow detection of single photons with $\alpha > 50$\%. The reduced capacitance also allows these devices to be extremely fast; typical photo-detector bandwidth is improved by a factor of 5-10 over conventional photomultiplier tube. The combination of these two improvements can allow an increase in $S/N$ of a factor of 3-6 over conventional photomultiplier tubes, and 2-4 over ultra-bialkali photomultiplier technology. 

Present technology SPADs are relatively inexpensive, and are stable with respect to intense light exposure, but suffer from comparatively large (80\,nsec) dead-time after observation of a light pulse. At the present time, a first implementation of a high speed, high quantum efficiency camera using SPAD technology has been constructed and tested.  This device, called AquEYE \cite{aqueye}, is a first generation implementation the ESO QuantEYE concept\cite{quanteye}. AquEYE employs four independent 50\,$\mu$m diameter SPAD pixels, with pixel timing better than 50\,psec and $\alpha >50$\% for $\lambda$ between 500\,nm and 600\,nm. AquEYE was first successfully tested in June 2007 on the 182\,cm Asagio Cima Ekar Telescope. Future technological development of SPAD type photo-detector technology will need to focus on reducing detector dead-time and overall system cost. 

\section{Activity, Organization, Partnerships and current status}
The main organizational body overseeing the development of Intensity Interferometry as a modern astronomic technique is the SII Working Group
within IAU commission 54 on optical and infrared interferometry. The SII
Working Group, commissioned in 2008, held its initial meeting  at the August
2008 SPIE meeting in Marseilles, France. At this meeting, it was decided to
organize an SII workshop in January 2009 at the University of Utah
\footnote{2009 workshop on Stellar Intensity
  Interferometry: http://www.physics.utah.edu/~lebohec/SIIWGWS/} in order to review the scientific and technical status of international SII development activities, and to coordinate research efforts.
Members of the SII working group are independently pursuing science topics and technical developments related to SII. Some of these activities include:
\begin{itemize}
\item Correlators, optical fibers, and photo-detectors are being developed and tested by independent research groups in the US and Europe.
\item The University of Utah is deploying a pair of 3\,m telescopes on a 20\,m
  East-West baseline \footnote{StarBase Utah:  http://www.physics.utah.edu/starbase/} in order to provide a realistic SII test environment for the SII working group.
\item SII Working Group Members of the VERITAS collaboration, in collaboration
  with Dravins, have performed inter-telescope correlation feasibility studies in October 2008 at the VERITAS Observatory. Correlations between fast intensity fluctuations between two VERITAS 12\,m diameter telescopes were studied  using the central photomultiplier tube of the VERITAS gamma-ray cameras.
 \item Close contacts are maintained with the CTA and AGIS collaboration regarding SII science topics and integration of SII specific electronics and secondary optics with the IACT gamma-ray cameras.
\item A working group focused specifically on SII science and SII camera
  integration, led by 
Dravins, has been established with the CTA collaboration. The CTA SII group has recently submitted a CTA Phase A design study for implementation of SII capability onto the CTA Observatory.
\item Theoretical studies of optimal image and phase reconstruction techniques are underway, with significant efforts in Israel, Utah, and Europe.
\end {itemize}
 Ongoing progress in all of these various directions necessitates frequent meetings to discuss results, ideas, and to plan new initiatives. The SII Working Group also serves as the main organization for development of the research collaboration that will develop the SII instrumentation and science research programs for the CTA and/or AGIS Observatories. We intend
to coordinate these various efforts through yearly meetings of the SII Working Group.

\section{Activity schedule}

The SII working Group envisions the development of the SII in four distinct Phases
\begin{enumerate}
\item Initial Prototype Phase (2008-2010)
\item IACT Array `Piggyback' Phase (2011-2013)
\item CTA/AGIS Science Program Phase (2013-2020)
\item Dedicated SII facility phase (2020-2030)
\end{enumerate}

During the Initial Prototype Phase (2008-2010), the SII working group is working
on parallel and complementary developments of experimental and
computational tools and facilities that will allow initial observations using
a modern SII system. Part of this work will involves using the Grantsville SII
3\,m telescopes to test out system components and reproduce measurements of
stellar diameters and binaries. This phase will generate initial
experience and validation of the SII technique which will prepare
the SII Working Group to deploy a multiple telescope ($>2$) SII system.

Experience and technology developed during Phase 1 will be refined and
standardized during Phase 2 which will likely include the instalation and
testing of a SII system at a Current IACT array.  The SII working group will
work with the VERITAS and HESS collaborations to develop a fully instrumented
SII system than can be deployed on either or both IACT arrays. The SII group
will submit science and development proposals to these observatories to deploy
and operate a full four telescope SII system. We anticipate performing 2 years
of science observations with the SII system during moonlight conditions,
accumulating up to 1600 hours
of source observations during this time.
During Phase 2,  an emphasis will also be placed on developing a reliable SII
system that can be fabricated and deployed on the larger CTA/AGIS arrays,
which should then be under construction. 

In 2013, we will enter Phase 3
 which includes operating a 50-100 telescope SII system. When the first CTA/AGIS telescopes are deployed, a small subset will be equipped with SII hardware for further testing (2013-2015) while the full scale implementation will be under preparation.
 By 2016, the entire CTA array will have been equipped with SII
hardware and the observation program will be carried until 2020. The large
number of redundant baselines sampled will substantially extend the
capabilities of the SII technique. It should be possible to establish a strong
science program with excellent image reconstruction and angular resolution in
the sub micro-arcsecond range.

Phase 4 would include the construction and operation of a dedicated SII $\rm
km^2$ area array of high signal bandwidth ($\rm\sim GHz$) telescopes. This would allow a substantial increase in the number of
available observation hours especially in dark sky (no moonlight)
conditions. This should result in increased sensitivity for observing more
distant (fainter) objects. This stage would take advantage of experience
gained with CTA/AGIS in the operation of large arrays of large light
collectors. This project is beyond the current decanal survey.

\section{Cost estimate}
As the proposed SII observations can be carried out using existing technology, the cost of many of the key items for the SII project can
already be reliably estimated. However, the basic design of the AGIS and CTA
telescopes is still under discussion, including the number of telescopes,
separations, diameters, etc. Consequently, the cost of the presently
envisioned full SII implementation on CTA/AGIS inherits cost estimate
uncertainties from uncertainties in the design of these observatories.
We can make a representative estimate of the SII implementation based upon a strawman CTA Observatory. The  strawman CTA Observatory employs 97 telescopes which will generate 4656 baselines. Each CTA
telescope will be equipped with secondary SII optics ($\sim \$5,000$) and at least two SII photo-detectors and associated power supply and slow control ($\sim
\$3,000$). SII signals will have to be communicated to the central station to be
processed. This communication will have to ensure synchronization to a
fraction of nanosecond in order to maintain $100\,MHz$ bandwidth. This can be achieved by optical fibers running to each telescope ($\$2,000$). The total cost for measuring the optical signal at each telescope and transporting the signals
to a central correlation/recording station is approximately $\$970,000$.
The two point correlation for each of the  4656 baselines can be achieved with
various techniques. Several correlator designs  are under study and all are
available for less than $\$200$ to which one would need to add the price of
the computing facility to centralize and record the data ($\$50,000$). The
full recording of the raw data at each telescope for off-line processing might
be possible for a similar or smaller cost. This brings the total hardware cost  of the entire SII add-on system to $\$1,951,200$.

Additional costs for construction and operation of the SII facility at CTA/AGIS include on-site manpower, travel, and infrastructure . We estimate
The SII construction phase will require at least two or three full time engineers/technicians on site over the three years construction phase. Typical
Cost for on-site personnel is estimated at \$600k for this construction phase. Infrastructure costs cannot realistically be estimated at this stage but can be expected to remain a relatively small fraction of the above mentioned costs.
Additional costs for student and faculty personnel have not been included above. We have not included costs for travel by off-site collaborators to the site to perform observations. Including a rough estimate of these costs, we estimate a
Total project cost (hardware, engineering, construction and operation over six year of Phase 3 operations) for an SII implementation at a single CTA (or AGIS ) IACT array observatory  of approximately \$5 million, including overhead.
This is less than 5\% of the projected CTA (or AGIS) construction budget.

\end{document}